\documentclass[aps,floats,twocolumn,showpacs]{revtex4}

\usepackage{pifont,multirow,graphicx,mathrsfs,makeidx,color,epsfig,
fancyhdr,floatflt,fancybox,calc,amsmath,amsfonts,amssymb,amsthm,latexsym}
\usepackage{epsfig}
\usepackage[latin1]{inputenc}

\newcommand{\nc}{\newcommand}
\nc{\be}{\begin{equation}} \nc{\ee}{\end{equation}}
\nc{\bea}{\begin{eqnarray}} \nc{\eea}{\end{eqnarray}}
\nc{\rds}{{\rm d}s} \nc{\rdt}{{\rm d}t} \nc{\rdr}{{\rm d}r}
\nc{\rdO}{{\rm d}\Omega} \nc{\s}{{\rm S}} \nc{\Pl}{{\rm P}}
\nc{\dis}{\displaystyle} \nc{\crit}{_{\rm cr}}
\nc{\rd}{{\rm d}} \nc{\munu}{{\mu\nu}}

\begin{document}

\title{ENERGY SPREAD OF THE UNSTABLE STATE\\ AND PROTON DECAY OBSERVATION}

\author{Giovanni Salesi}
\affiliation{Universit\`a di Bergamo, Facolt\`a di
Ingegneria, viale Marconi 5, 24044 Dalmine (BG), Italy
\\ and Istituto Nazionale di Fisica Nucleare, Sezione di Milano
via Celoria 16, I-20133 Milan, Italy}

\vspace*{1cm}

\begin{abstract}

\noindent Because of the extreme smallness of the energy spread of the unstable state describing the decaying proton, due in its turn to the anomalous smallness of the resonance width expected for the proton decay, the application of the Heisenberg time-energy relation predicts the measurement times for the proton decay observation to be so long as to forbid a ``continuous'' observation of the decay. This might account for the missing observation of the proton decay.

\end{abstract}



\maketitle

\section{Preliminary considerations}

\noindent In this letter we wish to put forth a simple conjecture qualitatively explaining the missing observation (up to now) of proton decay events. To this end, we shall employ the time-energy expression of the Heisenberg indetermination principle, following Ghirardi et al.\,\cite{A} and Fonda et al.\,\cite{B} (who, however, utilized it in relation to a different physical context concerning the decay observation: namely, the so-called Zeno effect with we shall not deal with here).

The time evolution of the (undisturbed) unstable microsystem may be described by the state vector
\begin{equation}
	\psi(t)=a(t)\psi_{0}+\psi_{\rm{decay}}(t),
\end{equation}
with $\left\langle \psi_{0}|\psi_{\rm{decay}}(t)\right\rangle=0$, where $a(t)$ is the non-decay amplitude, $\psi_{0}=\psi(0)$ is the initial so-called unstable state, and $\psi_{\rm{decay}}(t)$ accounts for the decay products; at the instant $t$ of observation of the decay state, we have the reduction $\psi(t)\rightarrow \psi_{0}$ if decay has not occurred or $\psi(t)\rightarrow\psi_{\rm{decay}}$ if decay has occurred.\\
Then the decay state observation, necessary for the experimental measurement of the lifetime, is made by a numerous series of measurements of yes-no type (decayed-not decayed), for each of which we can write the time-energy Heisenberg relation:
\begin{equation}
	T_{\rm{m}}\Delta E\gtrsim\hbar 
\end{equation}
where $T_{\rm{m}}$ is the duration of the measurement process by which it becomes possible to know whether the decay has or has not occurred, and $\Delta E$ is the energy mean fluctuaction in the state $\psi_{0}$. Let us observe that, if $\Delta E$ is sufficiently small (as expected for the extreme smallness of the resonance width $\gamma=\hbar/\tau$, $\tau$ being the proton lifetime), the relative $T_{\rm{m}}$ obtained by means of (2) is so large that single measurements cannot actually be executed in sequence during ordinary laboratory times: and this can substantially account for the missing observation of the proton decay. Unlike $\gamma$, whose numerical value is obtained, for example, by the Fermi golden rule, pratically depending only on the coupling constant of the force causing the decay, the energy spread $\Delta E$ is a physical quantity that may depend on the deepest details of the interaction Hamiltonian such as the cutoff function or on parameters relative to the experimental procedure employed during the measurement of the lifetime. The numerical value of $\Delta E$ does not affect the proton lifetime because the decay law $P(t)(=|a(t)|^{2})$ is always given by $e^{-\gamma t}$ except for very large times $t\gg\tau$ and for very small times $t\lesssim 10^{-23}$ s, the shape of $P(t)$ for these extreme times being strongly correlated to $\Delta E$.

In the following we shall examine two theoretical approaches employed for studying $P(t)$ at early times and that therefore give an estimate of $\Delta E$ sufficiently precise for our aims, since for small times we can assume $P(t)\simeq 1-(\Delta E)^2t^{2}/\hbar^{2}$. We shall see that in the first approach, essentially a perturbative procedure applied to the proton under observation as if it were isolated and undisturbed, only a peculiar choice of the cutoff in the energy distribution function can make the proton decay observable.

In the second approach, also accounting for the experimental devices utilized and for the disturbance introduced by the lifetime measuring apparatus, we finally realize that it is never possible (in reasonably short time) to know whether the proton is decayed or not: as a consequence it becomes impossible to measure its lifetime.
 
\section{Numerical evaluations}

\noindent Also with respect to the proton decay there have been proposed some theoretical procedures employing suitable propagators in the context of quantum field theory and time dependent perturbation theory. We essentially start from an interaction Hamiltonian of the following type,
\begin{equation}
H_{I}=-g\int\frac{\rd^3k}{(2\pi\hbar)^{3/2}}\frac{f(\omega)}{(2\omega)^{1/2}}
\psi_{D}^{\dagger}\psi_{d}^{\dagger}\psi_{p} + {\rm h.c.} ,
\end{equation}
where $d$ and $D$ indicate the light and the heavy decay product respectively, $\omega=\sqrt{k^2_d+m^2_d}$ is the light product's energy, $g$ is the coupling constant relative to the decay, and $f(\omega)$ is the cutoff function necessary for the convergence of the theory.

Even if, in so doing, we are not able to determine the unstable state $\psi_{0}$ and its energy spectrum $\rho(E)$ but only $P(t)$, we can (as the authors cited below do in the mentioned works) obtain from the small time expression of $P(t)\simeq 1-(\Delta E)^{2}t^{2}/\hbar^{2}$ a reliable estimate of $\Delta E$. Obviously, we must first of all fix the cutoff function $f(\omega)$. Sanchez--Gomez et al.,\,\cite{C} making for $f(\omega)$ the following choice,
\begin{equation}
	f(\omega)=\frac{M^{2}}{(M+\omega)^{2}}	\hspace{0.7cm} M={\rm O}(m_p)\approx 1\hspace{0.1cm} \rm{GeV}\,,
\end{equation}
get for $\Delta E$ an expression frequently employed when the energy spectrum $\rho(E)$ of the unstable state $\psi_{0}$ is almost symmetrical around the resonance energy:
\begin{equation}
	\Delta E\approx\sqrt{m_p-E_{\rm{th}}}\sqrt{\gamma}\hspace{0.3cm}(c=1),
\end{equation}
where $E_{\rm{th}}$ is the threshold energy for the decay products. For the important channel $p\rightarrow\pi^{0}e^{+}$, by assuming $E_{\rm{th}}$ equal to the sum of the rest masses of the final particles, from (5) and (2) we get
	\begin{subequations}
 \begin{equation}
 \Delta E\approx\sqrt{m_{p}-m_{{\pi}^{0}}-m_{{e}^+}}\sqrt{\gamma}\approx 28.3\,\sqrt{\gamma}\;\rm{MeV} 
 \end{equation}
 \begin{equation}
  T_{\rm{m}}\gtrsim\hbar/ \Delta E\approx 0.9 \times 10^{-12}\sqrt{\tau}\,\rm{s}
 \end{equation}
\end{subequations}
Miglietta and Rimini \cite{D}, very differently, making for the cutoff function the following choice,
\begin{eqnarray}
&& f(\omega) = \left\{
\begin{array}{ll}
1 & \quad \textrm{for} \ \omega < M \\
\ \\
0 & \quad \textrm{for} \ \omega > M \\ 
\end{array}
\right.
\label{bound}
\end{eqnarray}
with $M=\rm{O}(10^{15}$ GeV = grand-unification energy), get finally for the energy spread:
\begin{equation}
	\Delta E=\sqrt{M^{2}/4\rm{\pi}\Delta \textit{m}}\sqrt{\gamma},
\end{equation}
where $\Delta m$ is the mass difference between the final and initial particles. As $M\gg\Delta m$ ($M\approx 10^{14}\Delta m$), we have a numerical value of $\Delta E$ enormously larger than the previously reported value, and correspondingly we have a value of $T_{\rm{m}}$ much smaller. In the particular case here chosen, $p\rightarrow \pi^0e^+$, we shall get
$$
	\Delta E \approx 10^{16}\sqrt{\gamma}\hspace{0.1cm}\rm{MeV}
$$
\begin{equation}
	T_{\rm{m}}\gtrsim 2.6\times10^{-27}\sqrt{\tau}\hspace{0.1cm}\rm{s}
\end{equation}
Till now we have exclusively reviewed theoretical decay pictures, not accounting at all for the interaction occurring during the measurements between measuring apparatus and unstable microsystem, nor for the dependence, on the peculiar detector's parameters, of the energetic properties (as $\Delta E$) of the observed quantum state. There exist on the contrary dynamical models which take into account this, as those developed in the works by Watson and Goldberger \cite{E}, Bell and Goebel \cite{F} and mostly by Fonda and co-workers \cite{G,H,I,L,M}.

Since, as we said before, the decay state observation is made by a numerous series of measurements of yes-no type, and since such measurements are substantially equivalent to repeated localization operations, the above-mentioned authors have proposed unified treatment of the majority of the decays starting from the definition of $\psi_{0}$ as a wave function of two (or more) interacting and scattering particles localized within a fixed experimental distance $R$ from each other. Indeed, they showed that, if we have a resonance of width $\gamma$ in the cross section of two particles, the state vector $\psi_{0}$ that describes two particles well localized after the collision has an energy distribution function $\rho(E)$ that is Breit--Wigner-like: Thus corresponding to an unstable system being about to decay and that may therefore be assumed as the wave function of the initial unstable state. Generally, the decay products can be thought to be formed within a region of range $r\ll R$ (where $r$ is dependent only on the decay forces and $R$ only on the measurement apparatus) and, as long as they are far from each other at a distance $\lesssim R$, they are to be described by $\psi_{0}$; whereas they are able to interact with the surrounding environment and then to be observed by suitable measurements as decay products $(\psi_{\rm{decay}})$ only when they leave the region of range $R$ and become separately observable (only then the unstable particle really breaks!). To be able to make such a theoretical picture of the decay, the above-cited authors did utilize the quantum theory of scattering and the S-matrix properties. The explicit relative calculations can be found in the cited works and finally yield a $\rho(E)$ for the state vector $\psi_{0}$ well approximated by the following expression
\begin{eqnarray}
&& \rho(E) = \left\{
\begin{array}{ll}
\frac{\pi}{4\arctan(\hbar\,v)/R\gamma}\frac{\gamma/2\pi}{(E-m_p)^{2}+\gamma^{2}/4} & \quad \textrm{for} \ E \in \mathscr{E} \\
\ \\
0 & \ \ \ {\rm elsewhere} \\ 
\end{array}
\right.
\label{bound}
\end{eqnarray}
where $\mathscr{E} \equiv \left[m_p-\hbar v/2R, \ m_p+\hbar v/2R\right]$, \ $v$ is the relative speed of the two final particles, and $R$ is the experimental parameter, depending only on the measuring set-up, defined above. Correspondingly, for such an energy form factor, symmetrical around $E=m_{p}$ and going drastically to 0 for large $E$, we can well assume, as calculated directly by the authors:
$$
\Delta E\approx\sqrt{m_{p}-E_{\rm th}}\sqrt{\gamma}
$$
\begin{equation}
	=\sqrt{hv/2R}\sqrt{\gamma}
\end{equation}
We may assume (see refs. \cite{N} and \cite{G}) that $v\approx (10^{-2}\div 1)\,c$ and $R\approx (10^{-10}\div 10^{-8})$\,cm, relative to decays occurring directly inside counters, cloud, bubble or spark chambers, or observed by means of radiochemical methods. In the case of proton decay $v\approx c$, while $R$ may get a value up to $\approx 10$\,cm if the decay takes place in inert matter such as iron or cement (which the final particles have to cross before coming out and being revealed); actually, if we employ water together with the Cerenkov effect, $R$ is of the order of the molecular distances. Replacing such values in eq.\,(11), we have for $R\approx 10^{-8}\div 10\hspace{0.1cm}\rm{cm}$,
\begin{equation}
	\Delta E\approx 3\times10^{-2}\sqrt{\gamma}\div 10^{-6}\sqrt{\gamma}\,{\rm MeV}
\end{equation}
and then
\begin{equation}
	T_{m}\gtrsim 0.8\times10^{-9}\sqrt{\tau}\div 3\times 10^{-5}\sqrt{\tau}\,{\rm s}
\end{equation}

\section{Final conclusions}

\noindent Since, according to the most reliable predictions of the grand-unification theories, the proton lifetime should be $\gtrsim 10^{31}$ years, we have that $T_{\rm{m}}$ varies through all the theoretical models examined (except in the one proposed by Miglietta and Rimini) as follows
$$
	T_{m}\gtrsim 0.9\times 10^{-12}\sqrt{\tau}\div3\times 10^{-5}\sqrt{\tau}\,{\rm s}
$$
\begin{equation}
\approx (0.5\div1.7)\times 10^{7}\hspace{0.2cm}\,{\rm years}
\end{equation}
Very differently, in the model by Miglietta and Rimini we have
\be
	T_{m}\gtrsim 2.6\times 10^{-27}\sqrt{\tau}\,{\rm s} \approx 4.7\times 10^{-8}\,{\rm s}
\ee
The most reliable value of $(T_{\rm{m}})_{\rm{min}}$ should fall within the interval of values given by (14), and so is much larger than $4.7\times 10^{-8}$ s (actually such an extremely small value of the measurement duration is relative to a theoretical approach not accounting at all for the experimental procedure used). Moreover, we do recall that Fleming \cite{O} in 1983 argued that, if protons have to avoid a form of ``kinematical fragmentation'' before decaying, $\Delta E$ must be much smaller than eqs.\,(6) and (12) indicate: in such a way the lower limit for $T_{\rm{m}}$ would be even larger than predicted by eq.\,(14).

In conclusion, the uncertainty principle, because of the anomalous smallness of the energy spread due in its turn to the anomalous smallness of the resonance width $\gamma$, forbids the reduction $\psi(t)\rightarrow\psi_{0}$ or $\rightarrow\psi_{\rm{decay}}$ to occur in a reasonably short time during which a proton lifetime measurement can be performed. The result we have achieved reflects principally the difficulty of preparing or observing during short times a quantum state very close to an energy eigenstate, as the proton unstable state is expected to be.

\vspace*{1cm}

\noindent {\bf Acknowledgments}\\
\noindent We warmly thank M.\,Cini and E.\,Recami for useful and interesting discussion and suggestions and M.\,Longo for the very kind collaboration.

\end{document}